\documentclass[fleqn,usenatbib]{mnras}

\def\cm{{\rm\thinspace cm}}

\def\erg{{\rm\thinspace erg}}
\def\eV{{\rm\thinspace eV}}

\def\K{{\rm\thinspace K}}
\def\keV{{\rm\thinspace keV}}
\def\km{{\rm\thinspace km}}

\def\m{{\rm\thinspace m}}
\def\Mpc{{\rm\thinspace Mpc}}
\def\Msun{\hbox{$\rm\thinspace M_{\odot}$}}

\def\pc{{\rm\thinspace pc}}

\def\s{{\rm\thinspace s}}
\def\ks{{\rm\thinspace ks}}

\def\pcmcu{\hbox{$\cm^{-3}\,$}}

\def\ergpcmsqps{\hbox{$\erg\cm^{-2}\s^{-1}\,$}}

\def\ergps{\hbox{$\erg\s^{-1}\,$}}

\def\kmps{\hbox{$\km\s^{-1}\,$}}

\def\pcmsq{\hbox{$\cm^{-2}\,$}}
\def\pcmcu{\hbox{$\cm^{-3}\,$}}

\def\kmpspMpc{\hbox{$\kmps\Mpc^{-1}$}}

\def\h18{\hbox{H1821$+$643\,}}

\usepackage{newtxtext,newtxmath}

\usepackage[T1]{fontenc}
\usepackage{ae,aecompl}

\usepackage{graphicx}	

\title[Circumnuclear environment in NGC~1275]{Probing the circumnuclear environment of NGC~1275 with High-Resolution X-ray spectroscopy}

\author[C. S. Reynolds et al.]{
\parbox{16cm}{Christopher S. Reynolds$^{1}$\thanks{E-mail: csr12@ast.cam.ac.uk (CSR)},
Robyn N. Smith$^{2}$,
Andrew C. Fabian$^{1}$,
Yasushi Fukazawa$^3$,
Erin A. Kara$^4$, 
Richard F. Mushotzky$^{2}$,
Hirofumi Noda$^5$, 
Francesco Tombesi$^{6,7,2,8}$, 
and, Sylvain Veilleux$^{2}$}
\vspace{0.1cm}\\
$^{1}$Institute of Astronomy, University of Cambridge, Madingley Rise, Cambridge CB3 OHA, UK\\
$^{2}$Department of Astronomy and Joint Space Science Institute, University of Maryland, College Park, MD 20742, USA\\
$^3$Department of Physical Science, Hiroshima University, 1-3-1 Kagamiyama, Higashi-Hiroshima, Hiroshima 739-8526, Japan\\
$^4$MIT Kavli Institute for Astrophysics and Space Research, Cambridge, MA 02139, USA\\
$^5$Department of Earth and Space Science, Graduate School of Science, Osaka University, 1-1 Machikaneyama, Toyonaka, Osaka 560-0043, Japan\\
$^6$Department of Physics, Tor Vergata University of Rome, Via della Ricerca Scientifica 1, 00133 Rome, Italy\\
$^7$INAF-Osservatorio Astronomico di Roma, Via Frascati 33, 00078 Monte Porzio Catone, Italy\\
$^8$ NASA/Goddard Space Flight Center, Code 662, Greenbelt, MD 20771, USA
}

\date{Accepted XXX. Received YYY; in original form ZZZ}

\pubyear{2015}

\begin{document}
\label{firstpage}
\pagerange{\pageref{firstpage}--\pageref{lastpage}}
\maketitle

\begin{abstract}
NGC~1275 is the Brightest Cluster Galaxy (BCG) in the Perseus cluster
and hosts the active galactic nucleus (AGN) that is heating the
central 100\,kpc of the intracluster medium (ICM) atmosphere via a
regulated feedback loop.  Here we use a deep (490\,ks) Cycle-19 {\it
Chandra} High-Energy Transmission Grating (HETG) observation of
NGC~1275 to study the anatomy of this AGN.  The X-ray continuum is
adequately described by an unabsorbed power-law with photon index
$\Gamma\approx 1.9$, creating strong tension with the detected column
of molecular gas seen via HCN and HCO$^+$ line absorption against the
parsec-scale core/jet. This tension is resolved if we permit a
composite X-ray source; allowing a column of $N_H\sim 8\times
10^{22}\pcmsq$ to cover $\sim 15$\% of the X-ray emitter does produce
a significant improvement in the statistical quality of the spectral
fit.  We suggest that the dominant unabsorbed component corresponds to
the accretion disk corona, and the sub-dominant X-ray component is the
jet working surface and/or jet cocoon that is expanding into clumpy
molecular gas. We suggest that this may	be a common occurence in
BCG-AGN.  We conduct a search for photoionized absorbers/winds and
fail to detect such a component, ruling out columns and ionization
parameters often seen in many other Seyfert galaxies.  We detect the
6.4\,keV iron-K$\alpha$ fluorescence line seen previously by {\it
XMM-Newton} and {\it Hitomi}.  We describe an analysis methodology
which combines dispersive HETG spectra, non-dispersive
microcalorimeter spectra, and sensitive {\it XMM-Newton}/EPIC spectra
in order to constrain (sub)arcsec-scale extensions of the
iron-K$\alpha$ emission region.
\end{abstract}




\section{Introduction}\label{intro}

While being relatively rare in the Universe, cooling-core clusters of galaxies hold a special place in discussions of active galactic nucleus (AGN) feedback.  These are the systems in which the need for AGN feedback is most apparent; X-ray observations of the host intracluster medium (ICM) in such clusters find short radiative cooling times within the central 100\,kpc, yet there is an order of magnitude less cold gas and star formation in and around the central brightest cluster galaxy (BCG) than expected from such cooling \citep{peterson:06a,fabian:12a,liu:19a}.  There must clearly be a source of heat for the ICM core, and AGN heating is strongly implicated via observations of strong ICM/AGN-jet interactions including jet-blown cavities \citep{fabian:00a,heinz:02a,birzan:04a}, sound waves and weak shocks \citep{fabian:05a,graham:08a,million:10a}, and AGN-driven turbulence \citep{zhuravleva:14a,hitomi:16a}.  As such, cool core clusters are the ideal laboratory for understanding, at a detailed mechanistic level, at least one face of AGN feedback.

The process by which BCG AGN heat the ICM has been the subject of intense theoretical work and computational modelling \citep{churazov:01a,reynolds:02a,vernaleo:06a,li:14a,reynolds:15b,yang:16a,yang:16b,ruszkowski:17a,bambic:19a}.  Questions still remain about the dominant mode of heating (with weak shocks/sound waves, turbulent dissipation, and cosmic ray streaming as the primary contenders) but the basic ingredients of the heating problem now seem clear.  More mysterious is the other side of the feedback loop, the feeding of the black hole on sub-parsec scales in a manner that is self-regulated so as to produce the appropriate level of heating throughout the ICM core on the $>$10\,kpc scales. Broadly, the self-regulation of the feedback loop will be achieved if the AGN fuel supply results directly from the ICM cooling process, either by cooling-mediated Bondi-like accretion of the hot ICM \citep{allen:06a}, or the accretion of cold gas that condenses from the hot ICM due to thermal instability \citep{gaspari:13a}. Either way, we may expect an unusual mode of accretion to be operating for BCG AGN compared with typical AGN.  Thus, it is particularly interesting to examine the circumnuclear environment of BCG AGN, search for evidence of inflows/outflows that can be traced back to the ICM, and seek signs of any localized processes that may act to stabilize (or destabilize) the overall thermal regulation of the cluster core.

Here, we use high-resolution X-ray spectroscopy to examine the circumnuclear environment of the AGN in NGC~1275, the BCG at the centre of the Perseus cluster of galaxies. Perseus is the closest \cite[$z=0.0176$;][]{hitomi:16a} massive \citep[$M=6.6\pm 0.4\times 10^{14}\Msun$;][]{simionescu:11a} cool-core cluster and has become the Rosetta Stone of AGN-cluster feedback studies.   This gives studies of NGC~1275 special significance. While our investigation uses X-ray observations, it is important and relevant to review the findings from longer wavelength studies. Near infrared integral field unit data have revealed significant amounts of molecular hydrogen within 100\,pc of the black hole \citep{wilman:05a}. Some of the observed molecular material forms a rotating circumnuclear disk with radius $r\sim 50\pc$ \citep[][hereafter S13]{scharwachter:13a}, the dynamics of which imply a total enclosed mass of $M=8^{+7}_{-2}\times 10^8\Msun$. Using data from the {\it Gemini} Near-infrared Integral Field Spectrograph (NIFS), S13 argue that the observed H$_2$ lines are shock excited, and estimate that the circumnuclear disk contains $\sim 4\times 10^8\Msun$ of molecular gas orbiting the SMBH. They also find evidence for a streamer of molecular gas that appears to be inflowing towards this central disk in a retrograde sense, highlighting the kinematic complexity of this region. \cite{nagai:19a} present data from the Atacama Large Millimeter Array (ALMA) that traces the circumnuclear molecular gas with CO(2-1), HCN(3-2) and HCO$^+$(3-2) lines at high spatial resolution (20\,pc). They also find a complex set of molecular filaments and a cold rotating molecular disk extending 100\,pc from the SMBH, and highlight the similarity of this structure with predictions from the cold chaotic accretion model \citep{gaspari:17a}. The fact that the observed radio jet is oriented orthogonal to this molecular disk is suggestive that we are seeing the outer regions the SMBH accretion flow, and that this flow preserves its orientation down to close to the black hole. \cite{nagai:19a} discover HCN(3-2) and HCO$^+$(3-2) absorption against the radio continuum of the central pc-scale jet emission blueshifted by 300--600\kmps, suggesting a fast molecular outflow from the AGN with an estimated H$_2$ column density of $N_{H_2}\approx 2.3\times 10^{22}\pcmsq$. The existence of a fast molecular outflow on $R\sim 100\pc$ scales has been strengthened recently by new {\it Gemini} NIFS data that find a high velocity-dispersion component to the H$_2$ line which extends across the $3\arcsec\times 3\arcsec$ field of view of NIFS \citep{riffel:20a}.  

In this paper, we study the AGN and its circumnuclear environment using a deep (490\,ks) observation with the High-Energy Transmission Gratings (HETG) on the {\it Chandra X-ray Observatory}.  The 1--9\,keV band data is well described by a power-law\citep[modified by the expected Galactic absorption $N_H=1.32\times 10^{22}\pcmsq$][]{kalberla:05a}.  We recently used this fact to set the tightest limits to date on the coupling of photons and light axion-like-particles (ALP) in the magnetic field of the ICM \citep[][hereafter R20]{reynolds:20a}.  Here, we examine the implications of these data for the astrophysics of this AGN.  We quantify the absence of emission/absorption lines in the soft X-ray spectrum and show that NGC~1275 does not possess the ``warm absorber'' outflows that are typical of many Seyfert-like AGN. We do, however. find evidence that part (15--20\%) of the X-ray emission is subject to cold absorption ($N_H\sim 8\times 10^{22}\pcmsq$) suggesting a composite X-ray source and sub-parsec scale structure in the circumnuclear molecular gas.  We detect the iron-K$\alpha$ fluorescence line, previously seen by {\it XMM-Newton} and {\it Hitomi}, at high confidence. Motivated by hints of anomalous iron-K$\alpha$ broadening in the dispersed spectra, we discuss a methodology that can combined the strengths of the dispersive HETG data, non-dispersive microcalorimeter data from {\it Hitomi}, and high signal-to-noise but medium spectral resolution data from the {\it XMM-Newton}/EPIC in order to search for sub-arcsecond spatial extension of the iron-K$\alpha$ emission region.  

The paper is organised as follows.  Section~\ref{data} describes the {\it Chandra}/HETG observations that form our core dataset and sketches the reduction steps followed to produce science-ready products.   Section~\ref{broadbandspec} examines the broad-band HETG spectrum, reporting evidence for partial covering of the X-ray source by a substantial cold column.  Section~\ref{blind} reports a systematic ``blind'' search for emission and absorption lines in the 1.4--9\,keV spectrum, highlighting the importance of the look-elsewhere effect when assessing the significance of features. Section~\ref{outflows} presents a more global search for photoionized outflows using self-consistent photoionization models, that allow us to exclude the presence of absorbers with column densities and ionization parameters typically found in Seyfert nuclei. Section~\ref{reflection} examines the iron-K$\alpha$ fluorescence line and presents our methodology for constraining spatial broadening.  Our conclusions and the implications of our results are discussed in Section~\ref{discussion}.  Throughout this paper we adopt {\it Planck}-2018 cosmological parameters \citep[$H_0=67.4\kmpspMpc, \Omega_M=0.31, \Omega_\Lambda=0.69$; ][]{planck:18a}.  With a cosmological redshift to NGC~1275 of $z=0.0173$, this gives a luminosity distance of 78.0\,Mpc and a scale of 365\,pc per arc-second.  We note that, unless otherwise stated, all errors are quoted at the 90\% confidence level (CL). 
3
\section{The {\it Chandra}/HETG data reduction}\label{data}

\begin{table}
\caption{Log of the Cycle-19 {\it Chandra}/HETG observations of NGC~1275.}
\begin{center}
\begin{tabular}{cccc}
ObsID & Start date & Good exposure & Roll angle \\\hline\hline
& & (ks) & (deg) \\\hline
20823 & 2017-10-24 & 53.3 & $-45$ \\
20450 & 2017-10-27 & 29.6 & $-45$ \\
20826 & 2017-10-28 & 29.6 & $-49$ \\
20451 & 2017-10-30 & 36.5 & $-73$ \\
20837 & 2017-10-31 & 35.5& $-60$ \\
20838 & 2017-11-01 & 24.7 & $-60$\\
20839 & 2017-11-03 & 21.7 & $-66$\\
20840 & 2017-11-04 & 20.7 & $-70$\\
20449 & 2017-11-06 & 45.3 & $-129$\\
20841 & 2017-11-09 & 54.2 & $-154$\\
20842 & 2017-11-10 & 18.3 & $-143$\\
20843 & 2017-11-11 & 22.1 & $+120$\\
20824 & 2017-12-02 & 49.3 & $+117$\\
20827 & 2017-12-04 & 17.0 & $+111$\\
20844 & 2017-12-05 & 34.5 & $+111$\\\hline
\end{tabular}
\end{center}
\label{tab:obs}
\end{table}

The {\it Chandra}/HETG data that form the core of the present study have previously been described by R20.  Here, we use an updated but otherwise identical reduction of the data.    In brief, {\it Chandra} observed NGC~1275 as part of a Cycle-19 Large Project, with the observing being split into 15 segments (ObsIDs) between 2017 October 24 and 2017 December 5 (Table~\ref{tab:obs}). The HETG was placed in the X-ray beam resulting in two independent dispersed spectra, one from the High Energy Grating (HEG) and one from the Medium Energy Grating (MEG).  The HETG is a slit-less grating array resulting in dispersion of the extended ICM as well as the compact AGN in NGC~1275.  Still, as explicitly illustrated in Figure~1 of R20, the dispersed point source emission can be isolated from much of the background and dispersed ICM emission resulting in a high-quality and high-resolution spectrum of the AGN.

The raw data were reprocessed with CIAO-4.12 and CALDBv4.9.1.  The preparation of science-ready HEG and MEG spectra for the AGN then followed the standard recommended threads\footnote{http://cxc.cfa.harvard.edu/ciao/threads/spectra\_hetgacis/}, except for three modifications.  Firstly, the width of the extraction region was reduced by a factor of two from the default (i.e. we use {\tt width\_factor\_hetg=18}) in order to permit the extension of the HEG spectrum up to 9\,keV by reducing overlap of the MEG and HEG extraction regions at the center of the dispersion pattern.  Secondly, it was found that the automatic zeroth-order finder algorithm became confused by the extended ICM emission which, if uncorrected, would lead to a mis-identification of the center of the dispersed spectrum and hence an incorrect wavelength scale. This was overcome by hard-coding the initial guess for the location of the zeroth-order in the call to {\tt tgdetect} to be at the coordinates of the nucleus of NGC~1275 as determined by the spacecraft astrometry. We used sub-pixel imaging on a 0.05\arcsec grid to verify visually that, for each of the 15 ObsIDs, this leads to a correct placement of the center of the dispersed spectrum to within 0.2\arcsec or less.   We then extracted the (positive and negative) first order HEG and MEG spectra and background spectra for each of the 15 ObsIDs and computed the associated response matrices and effective area files.  The spectra were combined with the CIAO tool {\tt combine\_grating\_spectra} in order to produce a single first-order MEG and a single first-order HEG spectrum with a total of 490\,ks exposure time.  Higher (second and third) order spectra were produced and examined but do not have the signal-to-noise to be useful and hence will not be discussed further.

Thirdly, as discussed in R20, care was needed with the background subtraction of the AGN spectra.  The ``background'' for the AGN spectrum is principally due to dispersed (and some zeroth order) ICM light.  The ICM emission is centrally peaked around the AGN itself, but the background spectra must be extracted from offset strips; hence it is expected that the pipeline processing will produce background spectra that underestimate the true dispersed-ICM emission captured in the AGN extraction region.  Indeed, examining the default/pipeline products show that the background-subtracted AGN spectra still, actually, contain structure reminiscent of that in the background spectra. As shown in R20, MARX simulations find that this can be accounted for to a high accuracy with a simple renormalization of the background spectrum which we achieve by adjusting the AREASCAL keyword in the background spectra.  We find that the background features are formally minimized according to a C-statistic measure by setting AREASCAL to be 0.474 (HEG) and 0.525 (MEG), implying that the background spectra are scaled up in normalization by approximately a factor of two.  

\section{Broad-band HETG spectrum and the need for a composite X-ray source}\label{broadbandspec}

As shown in R20 (their Fig.~3), the HETG spectrum of NGC~1275 is well approximated by a power-law.  As a first look at the spectrum, we use C-statistic minimization to fit a power-law model simultaneously to the 1--7\,keV MEG and 1.5--9\,keV HEG spectra.  Cold/neutral Galactic absorption is included in the model fit with a column density of $N_H=1.32\times 10^{21}\pcmsq$ \citep{kalberla:05a}.  We allow a multiplicative offset when modeling two spectra to allow for absolute flux calibration offsets.  The best fitting model has a photon index $\Gamma=1.894\pm 0.009$, HEG power-law normalization at 1\,keV $A=(8.38\pm 0.08)\times 10^{-3}\,{\rm ph}\,{\rm s}^{-1}\,{\rm keV}^{-1}\,{\rm cm}^{-2}$, and a MEG offset factor of $1.026\pm 0.009$ (suggesting a slightly higher flux calibration for the MEG), with a goodness-of-fit of $C=4893$ for 4873 degrees of freedom. The corresponding 2--10\,keV flux is $(2.48\pm 0.02)\times 10^{-11}\ergpcmsqps$, and the 2--10\,keV (rest-frame) luminosity is $(1.67\pm 0.02)\times 10^{43}\ergps$. 

As discussed in the Introduction, ALMA has discovered HCN and HCO$^+$ absorption towards the parsec-scale jet \citep{nagai:19a} leading to an estimated H$_2$ column density of $N_{H_2}\approx 2.3\times 10^{22}\pcmsq$ (assuming an excitation temperature of 100K, an HCN-to-H$_2$ conversion factor of $10^{-9}$, and full covering of the mm-band continuum).  X-ray variability \citep{fabian:15b,imazato:21a} requires that a significant fraction of the observed X-ray continuum must originate from parsec scales or less, so it is interesting to ask whether there are any signatures of this molecular gas column in the X-ray spectrum.  If we assume that a cold (atomic$+$molecular) absorbing screen covers the entire X-ray source, the lack of soft X-ray spectral curvature leads to very tight limits on the column density --- adding a cold absorber at the redshift of NGC~1275 (xspec model {\tt zTBabs}) gives a formal limit of $N_H<3\times 10^{19}\pcmsq$ (90\% confidence level), three orders of magnitude smaller than the molecular column inferred from ALMA.  This difference between the X-ray and mm-band absorbing column densities appears robust to the underlying assumptions.  The difference is too great to be bridged by modifications of the assumed HCN excitation temperature or reasonable changes in the HCN-to-H$_2$ conversion factor.  On the X-ray side, a line-of-sight H$_2$ column of $\sim 2\times 10^{22}\pcmsq$ covering the whole source would be impossible to reconcile with the HETG data given any reasonable intrinsic spectrum (imparting a factor of 100 deviation from the observed power-law at 1\,keV).   The solution must be geometric, requiring the molecular gas to be clumpy or structured on sub-parsec scales.

\begin{figure}
\includegraphics[width=0.45\textwidth]{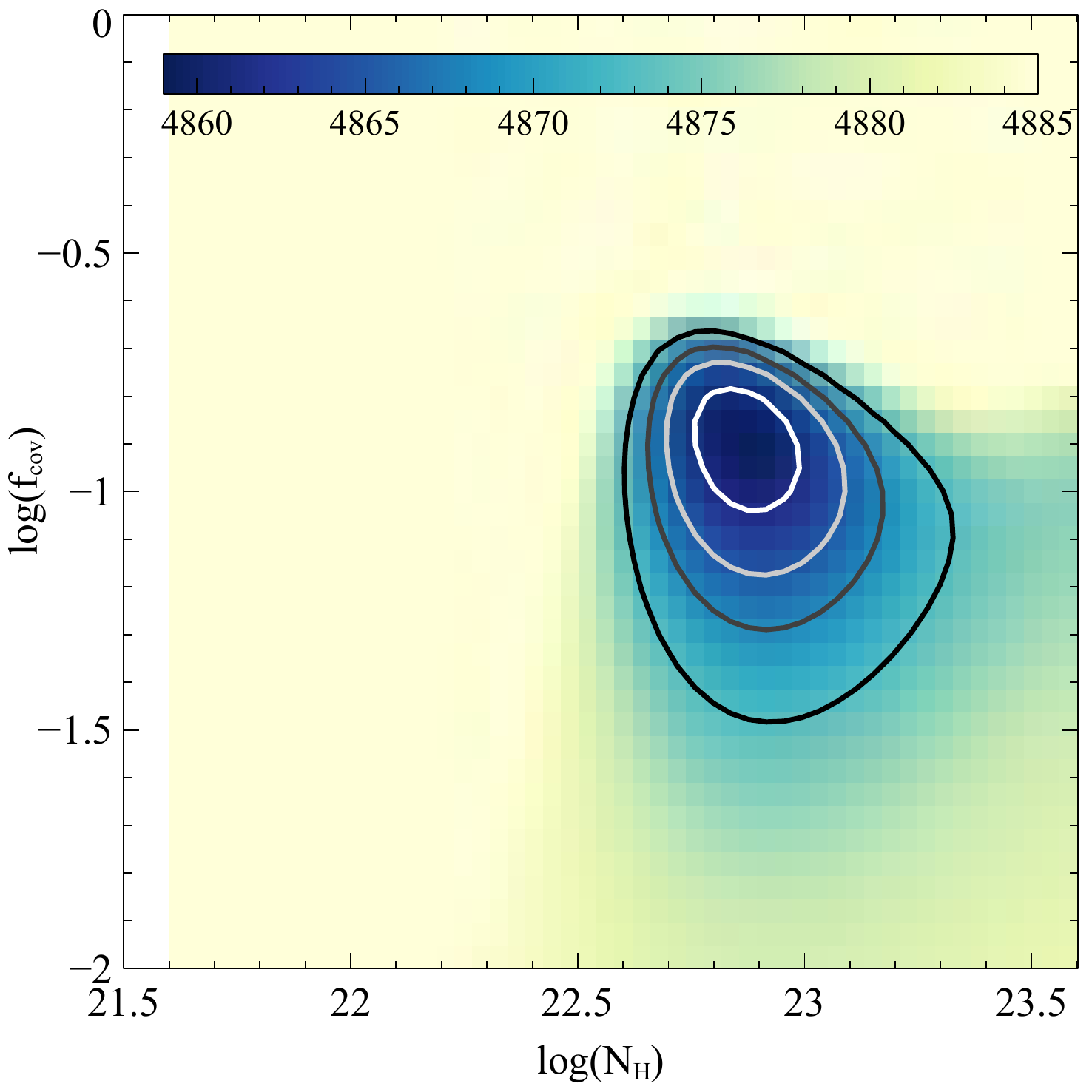}
\caption{Confidence contours on the $(\log N_H, \log f_{\rm cov})$-plane from the fit of a partially-covered power-law model to the HETG data.  Shown here the $68\%$ (white),$90\%$ (light grey), $99\%$ (dark grey) and $99.9\%$ (black) confidence levels, with the underlying C-stat in the background colour map.  }\label{fig:pcf_contours}
\end{figure}

The observed X-ray emission in NGC~1275 may be a composite such that it possesses absorbed and unabsorbed subcomponents, with possible contributions to the X-ray emission including the inner accretion disk, a sub-parsec scale jet core, and the hot-spot of the current jet activity which is currently 1.2\pc\ (projected) south of the nucleus \cite[radio component C3; ][]{nagai:10a,hodgson:21a}.   It is then possible that some component of the X-ray {\it does} experience the absorption from the molecular gas.  This motivates us to examine partial-covering solutions. The inclusion of a partial-covering absorber (xspec model {\tt TBpcf}) to our baseline power-law model leads to a significant improvement in the goodness of fit ($\Delta C=-24$ for two additional degrees of freedom) and flattens out some of the remaining ``wave-like'' residuals noted in R20.  In this pure HETG-fit, we infer an intrinsic column of $N_H=7.8^{+3.1}_{-1.9}\times 10^{22}\pcmsq$ covering a fraction $f_{\rm cov}=0.13\pm 0.04$ of the X-ray emission (Figure~\ref{fig:pcf_contours}). The inclusion of this partial-covering X-ray absorber steepens the best-fit power-law slope to $\Gamma=1.97\pm 0.03$.  This X-ray column density translates to a molecular column of $N_{H_2}=3.9^{+1.5}_{-1.0}\times 10^{22}\pcmsq$ which is approximately twice that inferred from {\it ALMA} assuming full covering of the mm-band continuum.  This suggests that the mm-band absorption itself may also composite.  We discuss this further in Section~\ref{discussion}

\section{A blind search for emission and absorption lines}\label{blind}

\begin{figure*}
\includegraphics[width=0.95\textwidth]{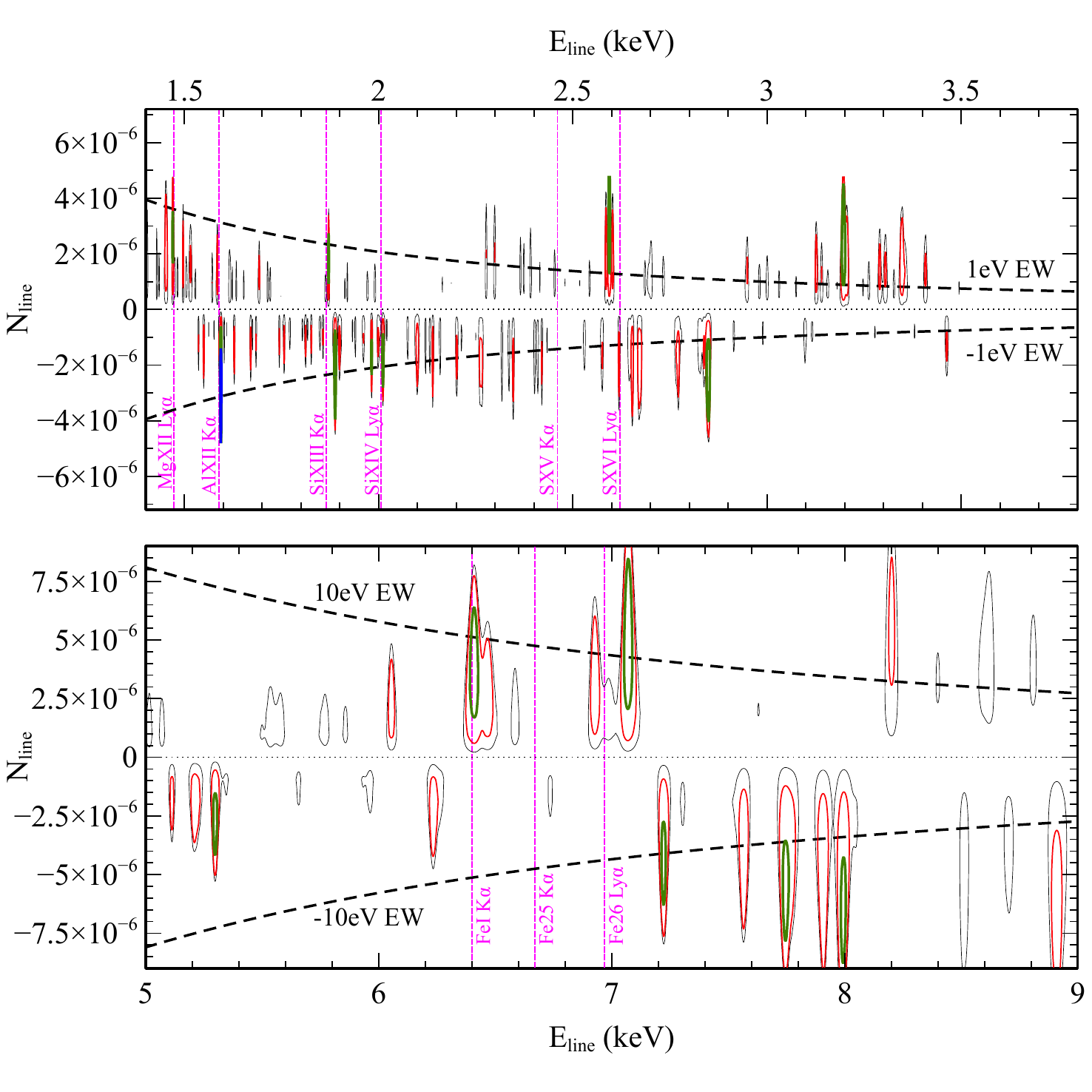}
\caption{Results of the blind line search in the MEG/HEG joint spectrum.  Shown are contours in the improvement in goodness of fit $\Delta C$ upon the inclusion of a Gaussian line with rest-frame energy $E_{\rm line}$ and normalization  $N_{\rm line}$ relative to a baseline model consisting of a simple power-law continuum subject to Galactic absorption ($N_{}\rm H)=1.32\times 10^{21}\pcmsq$.  Contour levels correspond to $\Delta C=-1.0$ (thin black line; 68\% single-trial CL), $\Delta C=-2.706$ red line; (90\% single-trial CL), $\Delta C=-6.65$ (green line; 99\% single-trial CL), $\Delta C=-13.8$ (blue line; 90\% CL for 500-trials).  Upper panel shows the results of the soft band analysis, with MEG and HEG fitted in the range 1--5\keV and 1.5--5\,keV respectively.  Lower panel shows the hard band analysis, with MEG and HEG fitted in the range 5--7\keV and 5--9\,keV respectively. }\label{fig:blindsearch}
\end{figure*}

These {\it Chandra}/HETG data permit the most sensitive search to date for X-ray emission and absorption lines from this AGN. Starting with the baseline power-law model discussed in Section~\ref{broadbandspec}, we examine the improvement in the C-statistic upon the inclusion of a Gaussian line with (rest frame) energy $E_{\rm line}$ and a normalization $N_{\rm line}$ that can be positive or negative.  The line is taken to be sufficiently narrow as to be unresolved by the HETG ($\sigma=1\eV$). We conduct the line search in two parts, treating the soft band (1.3--5\keV) and hard band (5--9\,keV) separately so that the results are not biased by any subtle curvature in the underlying continuum.  For this reason, this analysis is also insensitive to whether we use the simple power-law or the partial-covering model from Section~\ref{broadbandspec}; for simplicity, we use the simple power-law model.   For each of these two bands, we scan through a dense grid on the $(E_{\rm line}, N_{\rm line})$-plane, refit the continuum parameters, and record the change in C-statistic ($\Delta C$).

\begin{table*}
\caption{Results of fitting a narrow ($\sigma=1\eV$) Gaussian line to the most likely detections resulting from the blind line search of Section~\ref{blind}. }
\begin{center}
\begin{tabular}{cccclcl}\hline\hline
Energy 			& Norm 				& EW 			& $\Delta C$ & Possible ID 			& Velocity 		& Notes \\
(keV) 			& ($10^{-6}$ph/s/cm$^2$) & (eV) 			& 		& 						&($\kmps$)	&\\\hline
$1.469\pm 0.006$ 	& $2.5^{+2.0}_{-1.3}$	& $0.6\pm 0.4$	& $-8.4$ 	& MgXII Ly$\alpha$(1.472keV) 		& $+610\pm 220$			&\\
$1.593\pm 0.001$ 	& $-3.3\pm 0.1$		& $-0.9\pm 0.3$	& $-22.2$	& AlXII K$\alpha$(1.598keV)	& $+940\pm 110$			&\\
$1.871\pm 0.001$	& $1.8\pm 1.0$			& $0.7\pm 0.4$		& $-8.9$	& SiXIII K$\alpha$ (1.855keV)	& $-2590\pm 190$			& close to detector calibration feature\\
$2.011\pm 0.001$	& $-2.0\pm 1.0$		& $-0.8\pm 0.4$	& $-10.4$	& SiXIV Ly$\alpha$ (2.007keV)	& $-600\pm 190$			&\\
$2.619\pm 0.002$	& $-2.2\pm 1.4$		& $-1.5\pm 1.1$	& $-5.3$	& SXVI Ly$\alpha$ (2.620keV) 	& $+110\pm 230$			&\\
$6.411\pm 0.015$	& $3.7^{+2.2}_{-2.0}$	& $14\pm 8$		& $-9.43$	& FeI K$\alpha$ (6.40keV)	& $-520\pm 700$			& \\
$7.06^{-0.07}_{+0.12}$& $4.0^{+3.6}_{-2.1}$	& $18\pm 12$		& $-8.9$	& FeXXVI Ly$\alpha$ (6.97keV) & $-3870\pm 3000$			& \\\hline		
\end{tabular}
\end{center}
{Notes : All errors quoted at the  90\% level for one interesting parameter.  Line energy is given in the rest-frame of NGC~1275 assuming a redshift of $z=0.0173$. The improvement in the goodness of fit upon including the line ($\Delta C$) is with respect to the pure continuum model (powerlaw subject to Galactic absorption with $N_H=1.32\times 10^{21}\pcmsq$).  Velocity is quoted with respect to the rest-frame of NGC~1275, with positive velocity implying recession.}
\label{tab:lines}
\vspace{0.5cm}
\end{table*}%

Figure~\ref{fig:blindsearch} shows the results of this exercise. In the soft-band (upper panel), approximately 50 features exceed the 90\% single-trial confidence level (red contours), and eight exceed the 99\% single-trial confidence level (green contours). Four of these are close to, but not exactly at, energies of important transitions that might be expected from a photoionized AGN wind; MgXII\,Ly$\alpha$ (emission), AlXII\,K$\alpha$ (absorption), SiXIII\,K$\alpha$ (emission), and SiXIV\,Ly$\alpha$  (absorption). We also note a less significant feature (present at the 90\% but not 99\% single-trial CL) that is consistent with SXVI\,Ly$\alpha$. Table~\ref{tab:lines} reports the detailed properties of these tentative line detections.

\begin{figure}
\includegraphics[width=0.49\textwidth]{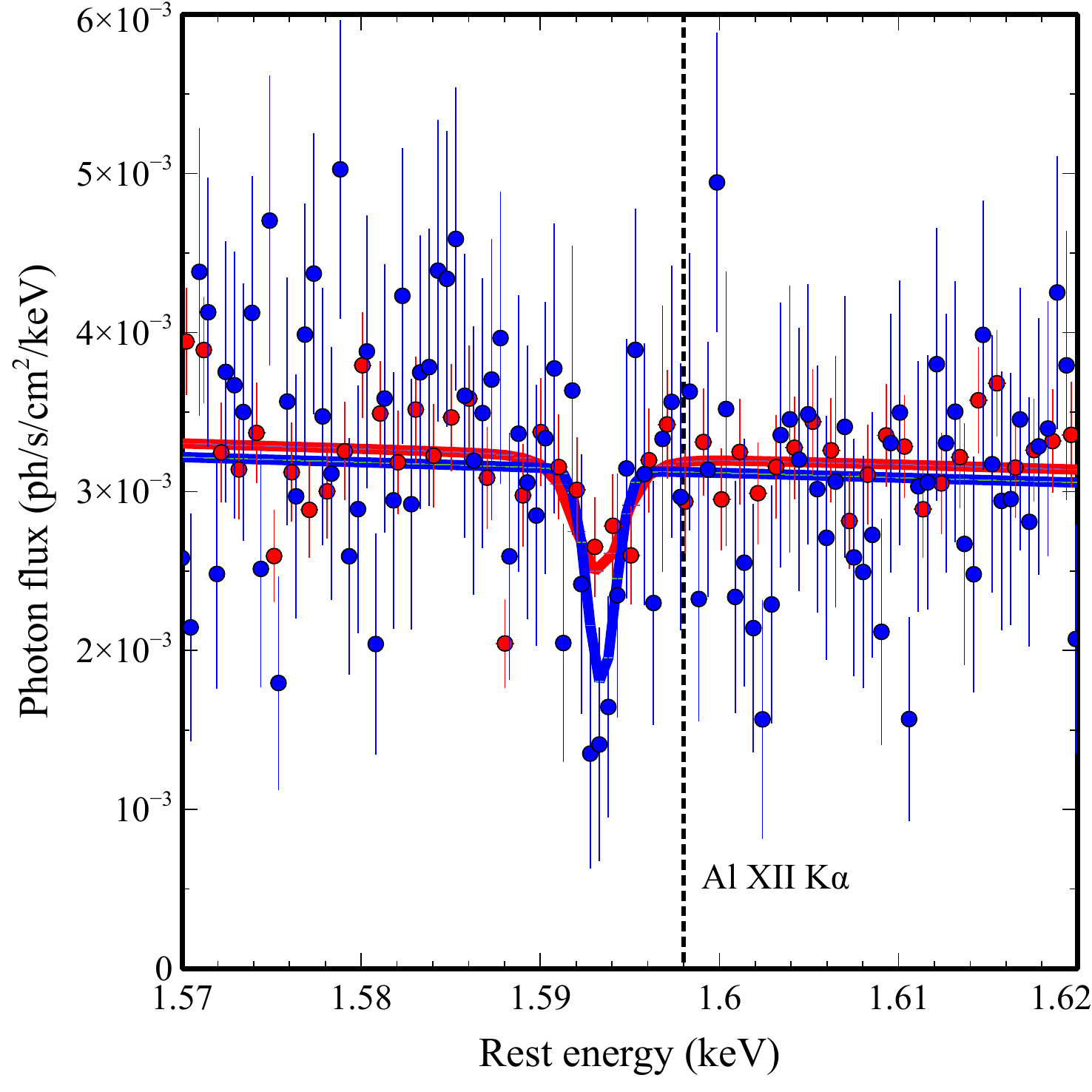}
\caption{Zoom-in of the 1.57--1.62\keV (rest-frame) region of the MEG (red) and HEG (blue) spectra showing the putative AlXII\,K$\alpha$ absorption line.  The solid red/blue lines show the corresponding best fitting power-law plus Gaussian model.  The laboratory rest frame energy of the AlXII\,K$\alpha$ line is shown as the vertical dotted line.}\label{fig:al12_line}
\end{figure}

When interpreting the features seen in this blind line search, care is needed; there are a large number of independent spectral bins within this band and so it is important to account for the look-elsewhere effect when assessing statistical significance. We take the fiducial soft-band resolution of the HETG ($E/\Delta E\sim 500$) as an estimate for the number of independent trials.  Only one feature, the putative AlXII\,K$\alpha$ feature at 1.593\,keV, exceeds the 90\% (500 trial) confidence level.  Figure~\ref{fig:al12_line} shows the AlXII\,K$\alpha$ region of the HEG and MEG with the best fitting Gaussian model.  

We note that, on the basis of the calculations by \cite{kockert:06a}, we do not expected to detect any absorption lines from the thermal ICM plasma. The strongest ICM absorption line in this soft band is predicted to be SXV\,K$\alpha$ (2.46\,keV) with an equivalent width (EW) of $\sim 1\eV$.  Such a weak line will fall below our sensitivity threshold and, indeed, we do not find such a feature.  The predicted EW of the ICM SiXIV\,Ly$\alpha$ and SXVI\,Ly$\alpha$ are approximately $0.2\eV$.  Thus, it seems highly unlikely that ICM absorption is responsible for the putative line detections reported in Table~\ref{tab:lines}.  An ICM origin for these lines is also inconsistent with the velocity shifts needed to bring the observed features in line with laboratory wavelengths.

Figure~\ref{fig:blindsearch} (bottom panel) shows the result of the blind line search in the hard-band.  We clearly detect the 6.4\,keV fluorescent K$\alpha$-line of cold iron at the 99\% (single trial) CL; this line is discussed much more extensively in Section~\ref{reflection}. We also detect, again at the 99\% (single trial) CL, an emission features that lies close to the FeXXVI\,Ly$\alpha$ line.  Making this identification suggests a significant flow of highly ionized plasma with a line of sight velocity of $V= -3900\pm 3000\kmps$ (negative sign indicating flow towards us relative to the rest-frame of the AGN).  There is a weaker and less significant feature ($E_{\rm line}=6.928\pm 0.018\keV$, $\Delta C=-5.6$) that could be a redshifted/counterflowing partner of this line with line of sight velocity of $V= +1700\pm 700\kmps$. This emission may result from a fast, recombining, highly-ionized wind driven by the central engine.  

Caution must again be exercised in interpreting any other features in this band.  Away from the physical energies for the astrophysically important lines, we must again take into account the look elsewhere effect.  We take the spectral resolution in this band ($E/\Delta E\sim 200$) as an estimate for the number of independent energy bins.  We find that none of the features in this band exceed a confidence level corresponding to 90\% for 200 trials and hence cannot be considered significant detections.

\section{Constraints on photoionized outflow signatures}\label{outflows}

The blind line  search reveals tentative detections of several absorption lines of high ionization species (AlXII\,K$\alpha$, SiXIII\,K$\alpha$, SiXIV\,Ly$\alpha$ and SXVI\,Ly$\alpha$) with blueshifts indicating velocities up to $\sim 2500\kmps$.  These may hint at the existence of a photoionized outflow from the AGN, similar to that found in half of all Seyfert nuclei \citep{reynolds:97b,tombesi:13a,laha:21a}. We also find a significant number of other absorption features further away from the rest-frame energies of expected strong transitions.  Especially interesting are the features in the 7--8\,keV band which are reminiscent of the highly ionized ultrafast outflows found in some Seyfert nuclei and broad-line radio galaxies.  While these lines are not individually significant given the look-elsewhere effect, it is interesting to note that pairs of these lines have separations close to the FeXXV~K$\alpha$/FeXXVI~Ly$\alpha$ separation raising the possibility that multi-species modeling might yield greater significance.

In this Section, we examine the HETG spectrum using self-consistent photoionization models.  We use the XSTAR photoionization code to compute the absorption spectrum from a column density $N_W$ of plasma photoionized by the AGN continuum. The ionization state of the plasma is described by an ionization parameter $\xi=4\pi F_{\rm ion}/n_e$ where $F_{\rm ion}$ is the energy flux of ionizing ($E>13.6\eV$) radiation incident on the plasma and $n_e$ is the electron number density.  For the purposes of the photoionization calculation, this ionizing continuum is modelled as a power-law with photon index $\Gamma=1.9$ between 0.1--100\,keV. We compute a grid of models that uniformly sample the $(\log\xi, \log N_W)$-plane with $20\times 20$ models spanning the range $\log\xi\in(0,6)$ and $\log N_W\in(20,24)$ (cgs units).  The resulting multiplicative absorption models can then be applied to any continuum model, passed through the instrumental response, and compared with the HETG data.

Guided by the results of Section~\ref{blind}, we model the full usable range of the MEG (1--7\,keV) and HEG (1.5--9\,keV) spectra with a baseline model consisting of a power-law continuum (modified by Galactic absorption) and narrow Gaussian emission lines at 6.4\,keV (fixed redshift $z=0.0173$) and 6.97\,keV (redshift free parameter).  Including the XSTAR photoionization model and minimizing CSTAT gives an improvement in the goodness of fit of $\Delta C=-11.8$, and suggests the presence of a highly ionized fast outflow with $\log\xi=3.92^{+0.37}_{-0.14}$, $\log N_W=22.3^{+0.17}_{-0.31}$ and an outflow velocity $v=(2.30^{+0.10}_{-0.05})\times 10^4\kmps$.  Here, the 90\% CL error bars are computed from the usual $\Delta C=2.706$ criterion.  However, as in the blind line search, great care must be taken in assessing the significance of any such outflow due to the unknown velocity giving rise to a strong look-elsewhere effect.  

\begin{table*}
\caption{Priors for our MCMC analysis of photoionized absorbers in NGC~1275. }
\begin{tabular}{llc}\hline\hline
Parameter & Prior & Range \\\hline
{\bf Photoionized absorber} & & \\
Column density $N_W$ 		& uniform in $\log N_W$ 	& $(10^{20},10^{24}\pcmsq)$ \\
Ionization parameter $\xi$ 	& uniform in $\log\xi$ 	& $(10,10^5\,{\rm erg}\,{\rm s}^{-1}\,{\rm cm})$ \\
Redshift (wrt to AGN) $z_{\rm ph}$ & uniform in $z_{\rm ph}$ & $(-0.3,0.1)$ \\\hline
{\bf Powerlaw continuum} & & \\
Photon index $\Gamma$ & uniform in $\Gamma$ & $(1,3)$ \\
Normalization@1\,keV $N_{\rm pl}$ & uniform in $N_{\rm pl}$ & $(0,10^{-1}\,{\rm ph}\,{\rm s}^{-1}\,{\rm cm}^{-2}\,{\rm keV}^{-1})$ \\\hline
{\bf Emission lines} & & \\
Normalization of 6.4\,keV line $N_{\rm 6.4}$ &  uniform in $N_{\rm 6.4}$ & $(0,10^{-4}\,{\rm ph}\,{\rm s}^{-1}\,{\rm cm}^{-2})$ \\
Normalization of 6.97\,keV line $N_{\rm 6.97}$ &  uniform in $N_{\rm 6.97}$ & $(0,10^{-4}\,{\rm ph}\,{\rm s}^{-1}\,{\rm cm}^{-2})$ \\
Redshift of 6.97\,keV line $z_{\rm 6.97}$ &  uniform in $z_{\rm 6.97}$ & $(0,0.05)$ \\\hline
MEG/HEG scaling ratio, ${\cal R}$ & uniform in ${\cal R}$ & (0.8,1.2) \\\hline\hline
\end{tabular}
\label{tab:mcmc}
\end{table*}

It is common practice to assess the significance of such detections in the presence of the look elsewhere effect by generating many simulated spectra with noise characteristics similar to the real data.  Instead, here we adopt a more Bayesian philosophy (i.e. one that avoids the invention of any data) and employ a Monte-Carlo Markov Chain (MCMC) analysis to map out the likelihood $P({\cal D}|{\cal  M})$ where $\cal{D}$ is the spectral dataset and ${\cal  M}=\{{\cal M}_i\}$ is the set of model parameters.  If we specify the prior probability for the set of model parameters $P(\cal{M})$, Bayes' theorem then allows us to calculate the basic quantity of interest, the posterior probability $P({\cal M}|{\cal D})\propto P(\cal{M})P({\cal D}|{\cal  M})$.  The probability distribution of any single parameter (or sub-set of ``interesting'' parameters) is then obtained by marginalizing $P({\cal M}|{\cal D})$ over the all uninteresting parameters.  Marginalization naturally accounts for any look-elsewhere effect that may be at play.

\begin{figure}
\includegraphics[width=0.48\textwidth]{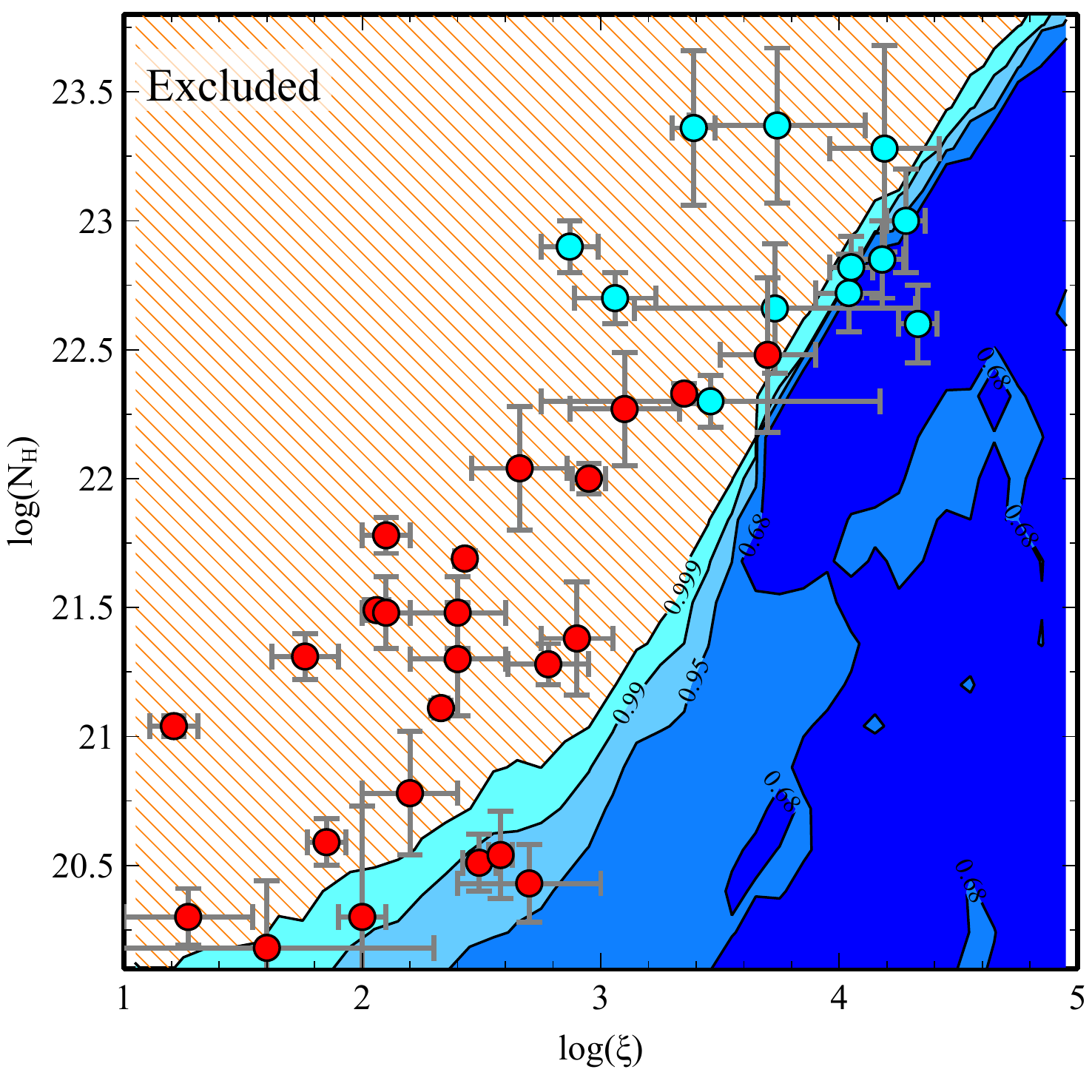}
\vspace{-0.5cm}
\caption{Constraints on the column densiity $N_W$ and ionization parameter $\xi$ of any photoionized absorber along the line of sight to X-ray source in NGC~1275.   Shown here are the 68\%, 95\%, 99\% and 99.9\% exclusion limits resulting from the MCMC analysis described in Section~\ref{outflows}.  For comparison, red datapoints show detected warm absorbers from the sample of Tombesi et al. (2013), and cyan datapoints show the detected ultrafast outflows from the sample of Tombesi et al. (2011).  These datapoints are shown with 1$\sigma$ errors. }
\label{fig:outflow}
\end{figure}

We use the Goodman-Weare MCMC algorithm incorporated within the XSPEC spectral fitting package to compute the likelihood for our problem.  The full set of model parameters and their relevant priors are listed in Table~\ref{tab:mcmc}.  We run eight independent MCMC chains each consisting of 100 walkers that undergo $1\times 10^4$ steps, giving chains with $1\times 10^6$ elements each.  Each chain has an initial burn-in consisting of $1\times 10^3$ steps (i.e. 10\% of the length of the full chain).   We find consistency between the chains in the distribution of all parameters, so we can combine them to give $8\times 10^6$ sample points for the likelihood and the corresponding posterior probability.

Figure~\ref{fig:outflow} shows the resulting joint probability distribution on the $(\log\xi, \log N_W)$ marginalizing over all other parameters.  We fail to detect a photoionized absorber, and exclude at high confidence the presence of an absorber across a large swath of the $(\log\xi, \log N_W)$-plane.  To give these  constraints context, Fig.~\ref{fig:outflow} shows the warm absorbers from the Seyfert sample of \cite{tombesi:13a} as well as the ultrafast outflows from the sample of \cite{tombesi:11a}.  We see that when photoionized absorbers are  detected, it is usually within the part of parameter space that we can already rule out for NGC~1275.  Thus the non-detection of a photoionized absorber in this object does indeed show our line of sight to be devoid of photoionized plasma columns that are commonly detected in bright AGN.

We note that if we repeat the MCMC analysis but freeze the outflow velocity of any absorber at $v=23,000\kmps$ (i.e. the value determined by simple minimization of CSTAT), then we would infer the presence of a high-ionization absorber with 97\% confidence.  This confirms that it is the look-elsewhere effect associated with the unknown velocity that compromises any such detection. 

\section{X-ray reflection from circumnuclear gas}\label{reflection}

Prior studies have already established the presence of the 6.4\,keV fluorescent K$\alpha$ line of cold iron in the X-ray spectrum of NGC~1275. This line was clearly detected in a 2001 observation by the {\it XMM-Newton}/EPIC with a reported equivalent width $\sim 165\eV$ \citep{churazov:2003hr}.  A follow-up {\it XMM-Newton}/EPIC in 2006 found the line with a similar flux, but the overall increase in the continuum flux led to the equivalent width decreasing to 70--80\,keV \citep{yamazaki:13a}. More recently, non-dispersive high-resolution spectroscopy with the Soft X-ray Spectrometer (SXS) on {\it Hitomi} also clearly identifies and, for the first time, resolves the line \citep{hitomi:18a} finding a full-width half maximum (FWHM) 500--1600\kmps  (90\% confidence level). Given this relatively low velocity width,  \cite{hitomi:18a} suggest that the line originates from a low-covering fraction atomic/molecular disk that extends anywhere from $\sim {\rm few}\pc$ to hundreds of pc from the black hole.  The flux of the line found by {\it  Hitomi} is consistent with that seen by {\it XMM-Newton}, entirely expected given that the light-crossing time of the fluorescent structure will be at least $\sim 10$ years.  


\subsection{Fluorescent iron emission in the HETG spectrum}

\begin{figure}
\includegraphics[width=0.48\textwidth]{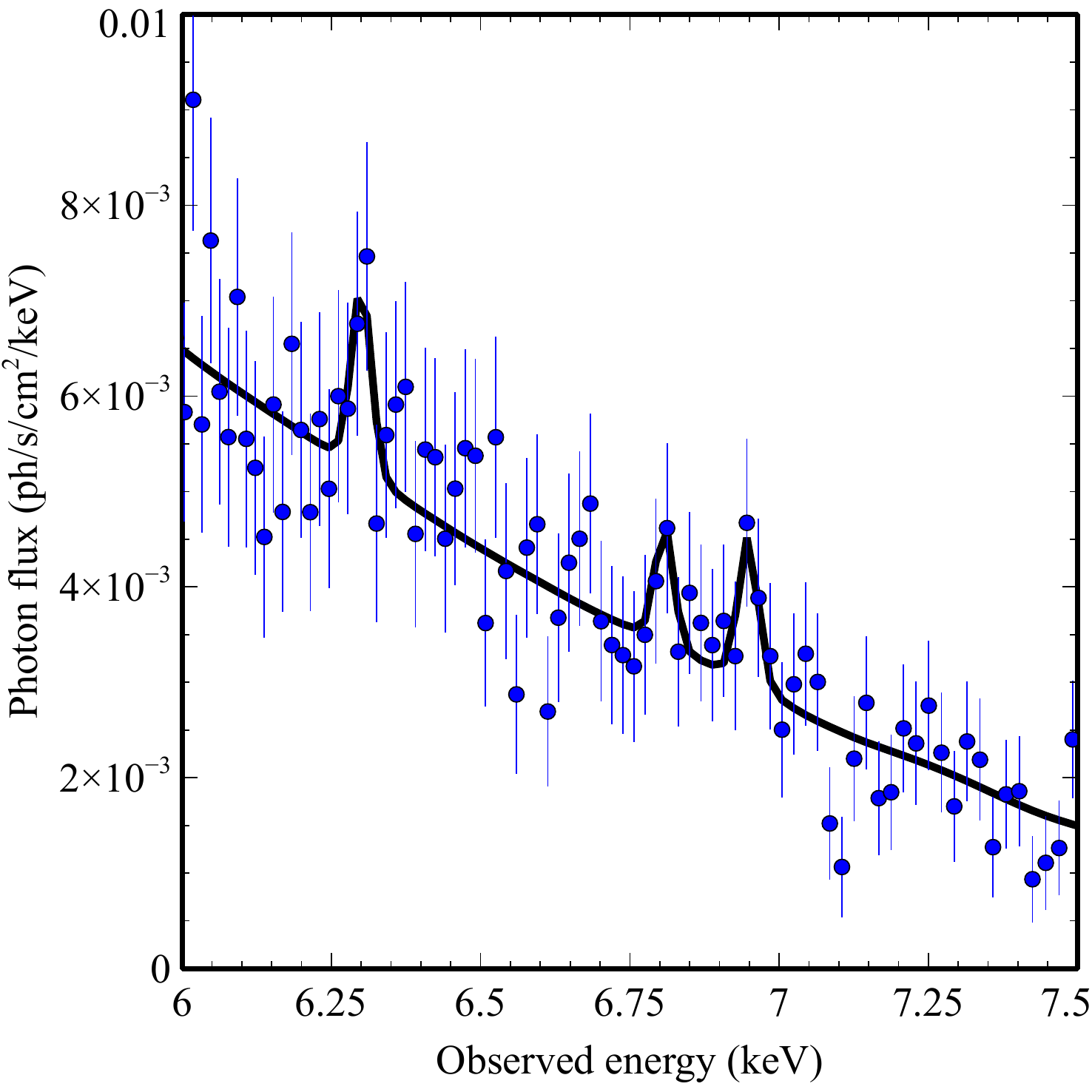}
\vspace{-0.5cm}
\caption{Iron K-band region of the HEG spectrum showing the neutral Fe-K$\alpha$ and the possible red/blue shifted FeXXVI-Ly$\alpha$ lines on top of the power-law continuum discussed in Section~\ref{blind}.  The spectrum has been lightly binned by a factor of two, resulting in oversampling of the spectral resolution by a factor of two.}
\label{fig:ironbandlines}
\end{figure}

As reported in Section~\ref{blind} and Table~\ref{tab:lines}, we also find the fluorescent iron-K$\alpha$ line in our HETG spectrum. Figure~\ref{fig:ironbandlines} shows the best fitting single narrow Gaussian model overplotted on part of HEG spectrum, lightly binned for plotting purposes.  To allow a more direct comparison with the {\it Hitomi} result, we proceed to refine our spectral model of the line to account for the fact that it is a doublet with energies 6.404\,keV (Fe-K$\alpha_1$) and 6.391\,keV (Fe-K$\alpha_2$) and a K$\alpha_1$/K$\alpha_2$ branching ratio of 2. Initially, we fix the redshift of the doublet to be strictly that of NGC~1275 ($z=0.0173$), and fix the velocity width of the emission line to be the best fitting value found by {\it Hitomi} ($\sigma=9\eV$). Jointly fitting to the HEG/MEG spectrum (in the 5--9\,keV/5--7\,keV bands, respectively) we find the Fe-K$\alpha_1$ normalization to be $2.6^{+1.6}_{-1.5}\times 10^{-6}\,{\rm ph}\,{\rm s}^{-1}\,{\rm cm}^{-2}$ (90\% CL). This is marginally in tension with the flux found by {\it Hitomi} ($4.5^{+1.5}_{-1.3}\times 10^{-6}\,{\rm ph}\,{\rm s}^{-1}\,{\rm cm}^{-2}$ at 90\% CL) and in somewhat stronger tension with reported {\it XMM-Newton} \cite[$8.4^{+4.1}_{-3.9}\times 10^{-6}\,{\rm ph}\,{\rm s}^{-1}\,{\rm cm}^{-2}$ at 90\% CL, dividing the total 2001 Fe-K$\alpha$ flux reported in][ by 1.5 to convert into a Fe-K$\alpha_1$ flux]{hitomi:18a}. 

Given the physical size of the fluorescing region implied by the {\it Hitomi} measurements of the line width, we consider a genuine change in line flux to be unlikely.  It is interesting to note that the HETG constraints on line flux can be brought into line with both {\it Hitomi} and {\it XMM-Newton} if we invoke additional line broadening.  Permitting the width of the Gaussian line model to be a free parameter in the HETG fits results in a very modest improvement in the goodness of fit ($\Delta C=-0.7$ with $\sigma<86\eV$ at 90\% CL) but this freedom increases the allowed range of Fe-K$\alpha_1$ line fluxes to $3.6^{+3.3}_{-2.2}\times 10^{-6}\,{\rm ph}\,{\rm s}^{-1}\,{\rm cm}^{-2}$ (90\% CL), entirely consistent with that found by {\it Hitomi} and {\it XMM-Newton}. 

\begin{figure*}
\includegraphics[width=0.45\textwidth]{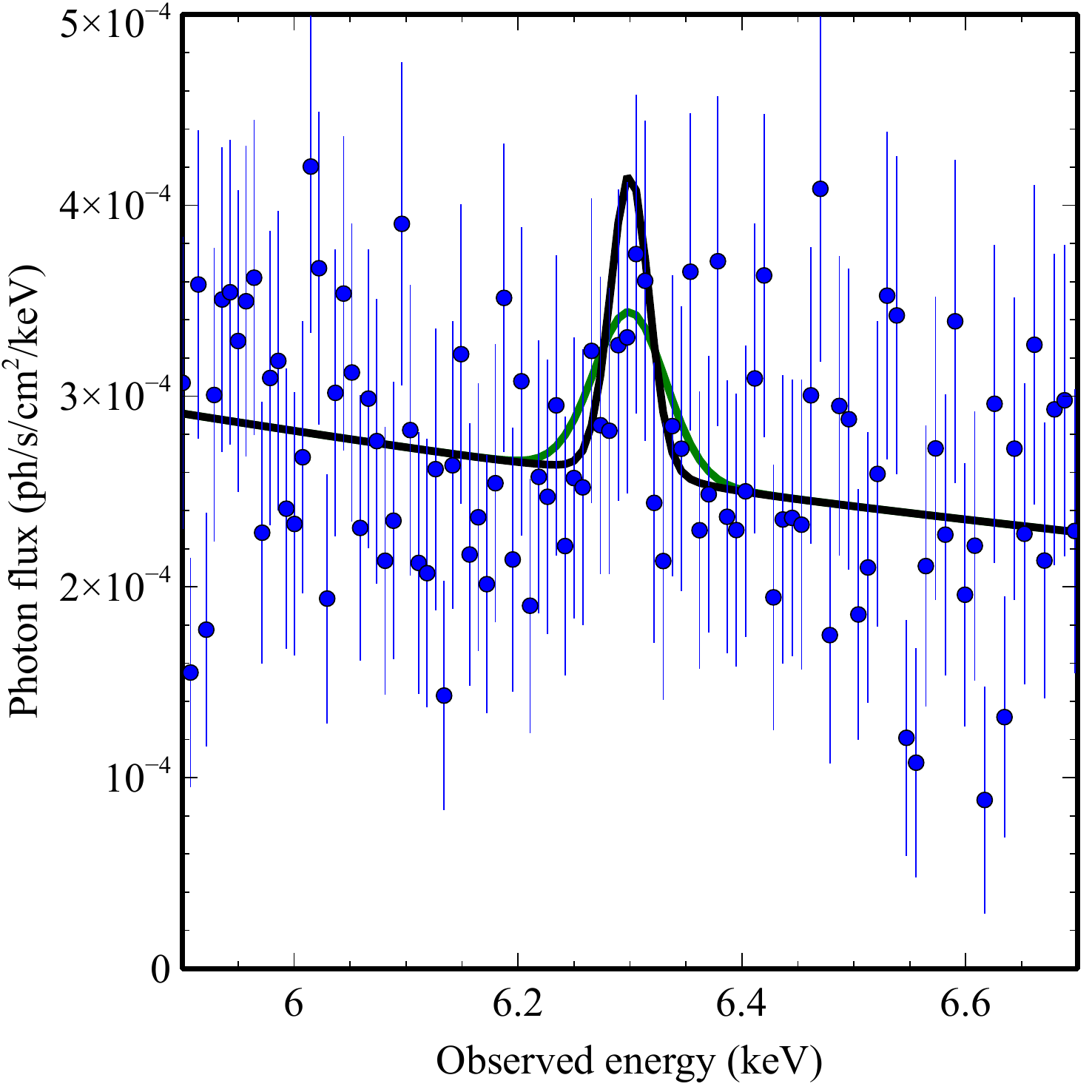}
\hspace{0.5cm}
\includegraphics[width=0.45\textwidth]{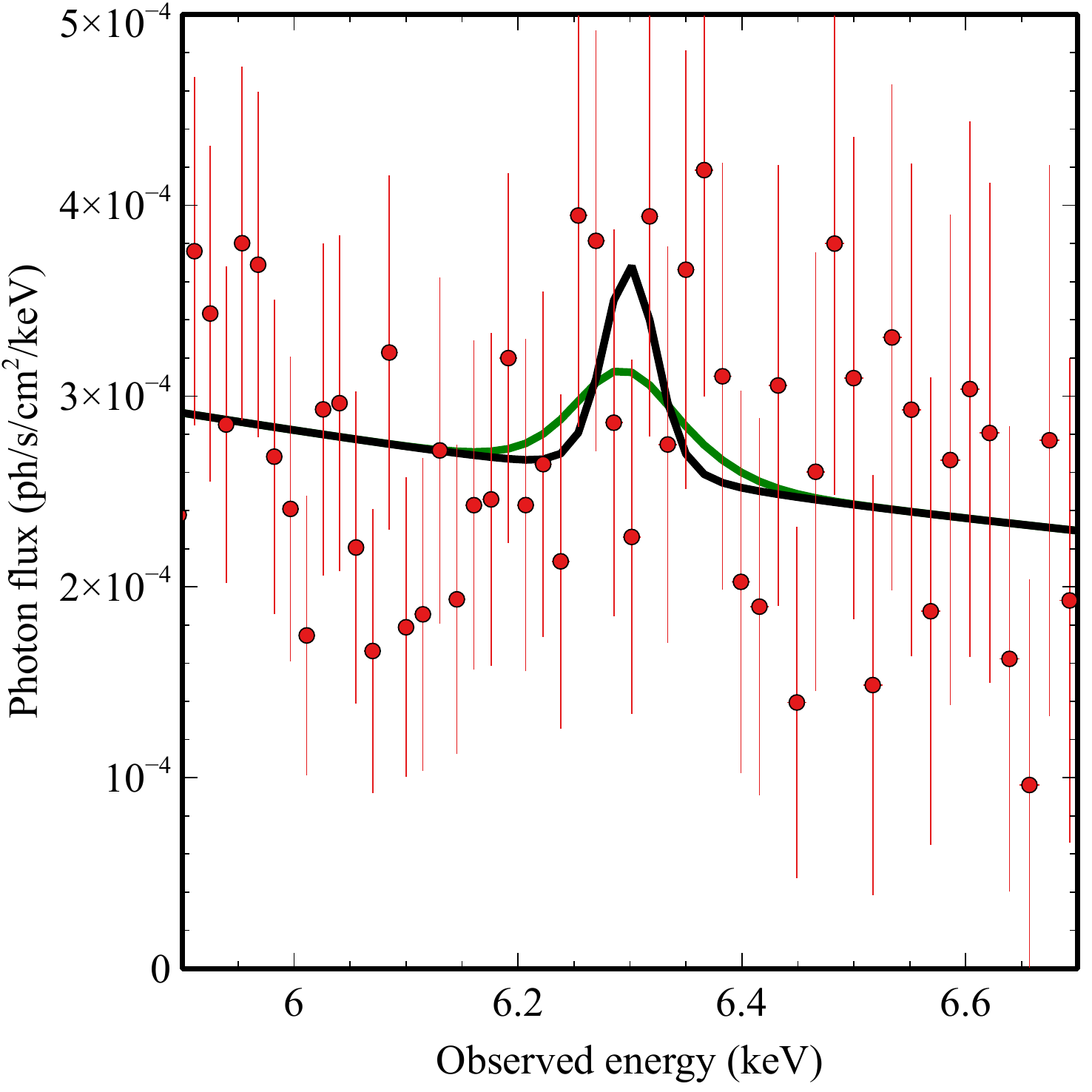}
\vspace{-0.2cm}
\caption{Illustration of the consequences of spatial extension on the inferred iron K$\alpha$ line in the unbinned HEG (left panel) and MEG (right panel) spectra.  The black line shows a model consisting of a point-like source of the iron-K$\alpha_1$/K$\alpha_2$ line with velocity dispersion $\sigma_v=420\kmps$ (corresponding to 9\,eV).  The green line shows the effect of extending the source of ron-K$\alpha$ photons by $\sigma_\theta=1\arcsec$. }
\label{fig:broadenedhetg}
\end{figure*}

The HETG is a slit-less dispersive spectrometer and so anomalous broadening of the Fe-K$\alpha$ line can result from spatial extension of the emitting gas \citep{marshall:17a}.  Assuming a Gaussian distribution of velocity (with dispersion $\sigma_{\rm v}$ in velocity units) and spatial radial profile (with standard deviation $\sigma_\theta$ in angular units), the overall measured line width $\sigma$ (in energy units) will be given by
\begin{equation}\label{eq:grating1}
\sigma^2=\left(\frac{E_0}{c}\right)^2\sigma_{\rm v}^2+\left(\frac{fFPE_0^2}{2\pi\hbar cmD_{\rm Row}}\right)^2\sigma_{\theta}^2,
\end{equation}
where $E_0$ is the rest-frame energy of the line centroid, $F=10.0548\m$ is the focal length of {\it Chandra}'s High Resolution Mirror Assembly, $P$ is the grating period ($P_{\rm MEG}=4001.95$\AA, $P_{\rm HEG}=2000.81$\AA), $m$ is the order of the dispersed spectrum, $D_{\rm Row}=8.63265\m$ is the Rowland distance of the HETG from the ACIS-S array, and $c$ and $\hbar$ are the usual fundamental constants.  We also include the correction factor of $f=0.73$ identified by Masterson \& Reynolds (in prep) on the basis of MARX simulations of extended iron-K$\alpha$ sources.  Evaluating eqn.~\ref{eq:grating1} for the first order ($m=1$) observations of the 6.4\,keV fluorescent lines gives HEG and MEG line widths of
\begin{equation}\label{eq:gratingheg}
\sigma_{\rm HEG}^2=45.5\left(\frac{\sigma_{\rm v}}{1000\kmps}\right)^2+7.37\times 10^2\left(\frac{\sigma_\theta}{1\,{\rm arcsec}}\right)^2\,\eV^2
\end{equation}
and 
\begin{equation}\label{eq:gratingmeg}
\sigma_{\rm MEG}^2=45.5\left(\frac{\sigma_{\rm v}}{1000\kmps}\right)^2+2.95\times 10^3\left(\frac{\sigma_\theta}{1\,{\rm arcsec}}\right)^2\,\eV^2
\end{equation}

As an illustration of this effect, Fig.~\ref{fig:broadenedhetg} shows the HEG and MEG spectra overlaid with two models for the iron line doublet, (i) a point-like source with velocity broadening corresponding to $\sigma_{\rm v}=9\eV$ corresponding to a FWHM of 990\kmps (black line), and (ii) a source with the same velocity broadening but spatially extended as a Gaussian radial profile with width $\sigma_\theta=1\arcsec$ (green line).  In both cases the normalization is set to be consistent with {\it Hitomi} and {\it XMM-Newton} (using the procedure outlined below).  It is apparent that the additional broadening due to spatial extension allows a higher flux line to be consistent with the HETG data.   We can also infer that, if the source were extended by too much, it would no longer be possible to fit the line seen most clearly in HEG. 

\subsection{Multi-observatory study of spatial extension}\label{reflectionmcmc}

Here, we present a methodology which allows rigorous constraints to be obtained on spatial broadening of the iron-K$\alpha$ fluorescing gas by demanding consistency in the properties of the emission line between {\it Chandra}/HETG, {\it Hitomi}/SXS and {\it XMM-Newton}/EPIC observations.  While we ultimate conclude that extension of the iron-K$\alpha$ source is not required in NGC~1275 at the 90\% confidence level, we describe this analysis in detail as we believe that it may find application to future datasets once  additional microcalorimeter observations of AGN are made by the {\it X-ray Imaging Spectroscopy Mission (XRISM)}.  

We perform a joint analysis of the  {\it Chandra}/HETG, {\it Hitomi}/SXS and {\it XMM-Newton}/EPIC data. For the {\it Chandra}/HETG, we consider the 1--7\,keV MEG and 1.5--9\,keV HEG spectra already described.  For {\it Hitomi}, we use the non-dispersive high-resolution micro-calorimeter SXS spectrum of the core of the Perseus cluster spectrum previously described by \cite{hitomi:18a}.  This spectrum will drive our constraints on the true velocity broadening of the fluorescent iron line independently of any spatial extent.  To recap the  {\it Hitomi} reduction, we use the {\it Hitomi} observations of the Perseus cluster obtained on 25--27 February 2016 and 4--6 March 2016 (OBSIDs 100040020--100040050) giving a total on-source exposure time of 240\,ksec. The cleaning of the events list, spectral calibration, and characterization of the non-cosmic background is discussed in detail in \cite{hitomi:18a}.  From the cleaned events list, we extract a spectrum of the $3\times 3$ pixel$^2$ region centred on NGC~1275; this choice of region is estimated to capture at least 95\% of the AGN flux.  We employ the spectral response matrix that includes the ``parabolic correction'', and the effective area curve appropriate for a point-like source.  

We also include the {\it XMM-Newton} observations of NGC~1275 taken 30--31 January 2001; the {\it XMM-Newton} data provide the most stringent constraints on the flux of the fluorescent iron line.  After obtaining the data from the {\it XMM-Newton} Science Archive (XSA) archives, we reprocessed the data using the {\it XMM-Newton} Science Analysis Software (SAS) version 18.0.0 and the Current Calibration Files (CCF) as of 28-June-2020.  We then followed the standard data analysis threads\footnote{https://www.cosmos.esa.int/web/xmm-newton/sas-threads} in order to clean the events files of background flares and extract EPIC-MOS and EPIC-pn spectra of NGC~1275.  A preliminary AGN spectrum was extracted from a circular region of radius 15\,arcsec centred on the core of NGC~1275, and a background spectrum (which is dominated by ICM emission) is extracted from an annulus centred on the AGN with inner--outer radii of 15--30\arcsec.  An examination of the single-to-double event ratio suggests that the EPIC spectra are affected by modest photon pile-up.  We mitigate the effects of pileup by excising counts from the central 3\arcsec (radius).  An additional {\it XMM-Newton} observation taken on 29--31 January 2006 was examined, but the source was brighter thereby increasing the effects of pileup to the extent that we decided not to use these data. 

We perform a joint fit of six datasets spanning three epochs; a single {\it Hitomi}/SXS spectrum (3--10\,keV band), the three {\it XMM-Newton}EPIC-MOS1/MOS2/pn spectra (2-10\,keV band), and the two {\it Chandra}/HETG spectra (using 1.5--9\,keV for HEG, 1--7\,keV for MEG). We use a spectral model consisting of an AGN power-law continuum subject to a partially-covering cold absorber, the fluorescent iron-K$\alpha_1$/K$\alpha_2$ doublet, and optical thin thermal emission from the plasma of the ICM core, all subject to neutral Galactic absorption with a column density $N_H=1.32\times 10^{21}\pcmsq$.  The ICM emission is described with the APEC model that includes thermal and turbulent line broadening (XSPEC model {\tt bapec}); the explicit inclusion of ICM line broadening is crucial given the inclusion of the high resolution {\it Hitomi}/SXS data. Both the photon index and normalization of the AGN power-law continuum are allowed to be free parameters for each of the three epochs.  The ICM component needs to be treated with more care in the joint fit of this rather heterogeneous dataset.  For the {\it Hitomi} observation, the spectrum includes the AGN and a dominant contribution from the ICM; we freely fit for the plasma temperature $T$, metallicity $Z$ \citep[with respect  to cosmic values of ][]{anders:89a}, ICM redshift $z_{\rm icm}$, velocity dispersion $\sigma_{\rm icm}$, and normalization.  Due to the smaller extraction region and ability to define an appropriate background region, the {\it XMM-Newton} spectra have a sub-dominant thermal plasma component from the centralmost regions of the cluster. Thus, in these spectra, the ICM temperature and normalization for the block of {\it XMM-Newton} are free parameters, but the metallicity, redshift and velocity dispersion are tied to the {\it Hitomi} model.  The {\it Chandra}/HETG spectra, given the background renormalization described in Section~\ref{data}, have no contribution from ICM emission and hence that component is given zero normalization when fitting the HEG and MEG spectra.

Given that the AGN emission is sub-dominant in the {\it Hitomi}/SXS spectrum, it is not possible for these data to provide meaningful constraints on the absorber that partially-covers the AGN X-ray continuum.  We choose to tie the absorber parameters in this epoch to those of the HETG spectra which were taken approximately 20 months later.  We do, however, allow the parameters of this absorption to be different at the time of the {\it XMM-Newton}/EPIC observation,15 years earlier than the {\it Hitomi}/SXS campaign.  

For the fluorescent iron line, we demand that the flux and velocity width $\sigma_v$ of the iron-K$\alpha_1$/K$\alpha_2$ doublet (with rest-frame energy 6.404\,keV/6.391\,keV and branching ratio 2:1 respectively) is fixed across all datasets. Additional broadening of the line in the dispersed {\it Chandra}/HETG spectra due to spatial extent $\sigma_\theta$ is including according to eqns.~\ref{eq:gratingheg} and \ref{eq:gratingmeg}.  

We again follow a Bayesian approach, using the Goodman-Weare MCMC algorithm to map out the likelihood through the 23-dimensional parameter space defining this model.  Priors for the parameters are listed in Table~\ref{tab:jointfitting}.  We use 640 walkers to span the parameter space, and perform an MCMC run with a burn-in length of $6.4\times 10^5$ (1000 steps for each of the walkers), and a main chain with length $1.28\times 10^6$ (2000 steps for each walker).  Likelihood is assessed using the total CSTAT for the model fit to the six spectra.  The {\it Hitomi}/SXS and {\it Chandra}/HETG spectra are unbinned apart from the elimination of energy channels with zero counts.  For computational expediency, the {\it XMM-Newton}/EPIC spectra are binned to a minimum of 25 photons per bin.  

\begin{table*}
\caption{Results of MCMC analysis of the iron-K$\alpha$ line from the joint {\it XMM-Newton}/{\it Hitomi}/{\it Chandra}.  All error are quoted at the 90\% CL.}
\begin{tabular}{llcccc}\hline\hline
Parameter 			& Prior 						& {\it XMM-Newton} 		& {\it Hitomi}/SXS 		& {\it Chandra}/HETG \\
					&							&(Jan-2001)			& (Feb/Mar-2016)		& (Oct/Nov-2017) \\\hline
{\bf AGN power-law} 		& & & & \\

Photon index $\Gamma$	& uniform in $\Gamma\in(0,3)$		& $1.65^{+0.03}_{-0.04}$		& $2.12^{+0.10}_{-0.11}$ 	& $2.03^{+0.03}_{-0.04}$ \\
Normalization $A$		& uniform in $A\in (0,10^{-1})$ 		& $(2.4^{+0.6}_{-0.3})\times 10^{-3}$ & $(1.4^{+0.5}_{-0.4})\times 10^{-2}$ & $(1.1\pm 0.1)\times 10^{-2}$ \\
{\bf X-ray partial covering cold absorber} &  & & & \\
Column Density $N_H$ ($\pcmsq$)	& uniform in $N_H\in (0,10^{24})$ & 	$(5.7^{+4.5}_{-1.6})\times 10^{23}$ & ={\it Chandra}& $(8.2^{+2.5}_{-1.3})\times 10^{22}$\\
Covering fraction $f_{\rm cov}$ & uniform in $f_{\rm cov}\in (0,1)$ & $0.37^{+0.14}_{-0.09}$ & ={\it Chandra} & $0.19\pm 0.04$\\
\hline
{\bf Thermal ICM}		&  & & & \\
Temperature $T$ (keV)	& uniform in $T\in (10^{-2},64)$		& $0.82^{+0.05}_{-0.06}$	& $3.53^{+0.09}_{+0.08}$ 	& ---\\
Metallicity $Z$ ($Z_\odot$)& uniform in $Z\in (0,5)$ 			& ={\it Hitomi}			& $0.49^{+0.07}_{-0.05}$ 				& ---\\
Redshift $z_{\rm icm}$	& uniform in $z_{\rm icm}\in (0,0.1)$ 	& ={\it Hitomi} 			& $0.01777^{+0.00001}_{-0.00003}$		& ---\\
Velocity dispersion $\sigma_{\rm icm}$ (\kmps)& uniform in $\sigma_{\rm icm}\in (0,10^4)$ & ={\it Hitomi} & $180^{+7}_{-8}$		& ---\\
Normalization $A_{\rm icm}$ & uniform in $A_{\rm icm}\in (0,10^{-1})$ & $(3.0^{+0.6}_{-0.5})\times 10^{-4}$ & $(5.9^{+0.6}_{-0.7})\times 10^{-2}$ 	& ---	\\\hline
{\bf Iron-K$\alpha$ emission} & & & & & \\
Redshift $z_{\rm Fe}$	& uniform in $z_{\rm Fe}\in (0,0.1)$ & ={\it Hitomi} 			& $0.0176^{+0.0008}_{-0.0007}$	& $0.011^{+0.009}_{-0.028}$\\
Velocity dispersion $\sigma_v$ (\kmps)& uniform in $\sigma_v\in (0,10^4)$ & \multicolumn{3}{c}{$424^{+539}_{-114}$ (tied across all datasets)}\\
Spatial extent $\log\sigma_\theta$ (arcsec)& uniform in $\log\sigma_\theta\in (-2,1)$ & \multicolumn{3}{c}{$0.1^{+0.8}_{-1.5}$ (tied across all datasets)}\\
K$\alpha_1$ normalization $A_{\rm Fe}$ & uniform in $A_{\rm Fe}\in (0,10^{-1})$ & \multicolumn{3}{c}{$(4.5^{+1.4}_{-0.9})\times 10^{-6}$ (tied across all datasets)}\\
K$\alpha_1$+K$\alpha_1$ equiv. width $W_{K\alpha}$ (eV) & (derived) 	& $77^{+24}_{-15}$ 	& $24^{+7}_{-5}$ & $27^{+9}_{-6}$  \\\hline
2--10\,keV AGN flux ($10^{-11}$ cgs) & (derived; absorbed flux) 	& 0.74 & 2.8 & 2.5 \\
2--10\,keV AGN luminosity ($10^{43}$ cgs) & (derived; de-absorbed)	& 0.71 & 2.2 & 2.0\\\hline\hline
\end{tabular}\label{tab:jointfitting}
\end{table*}

\begin{figure}
\includegraphics[width=0.45\textwidth]{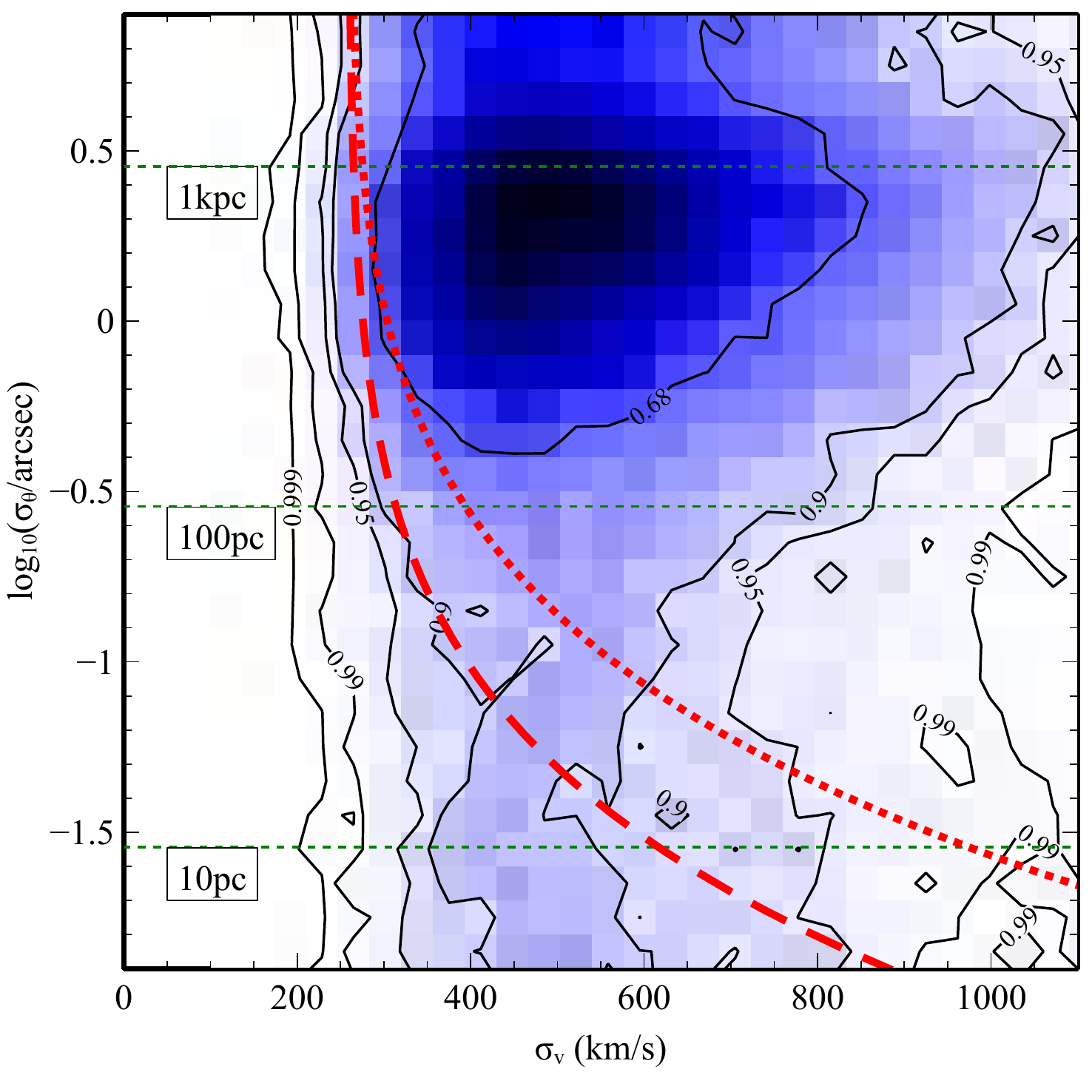}
\caption{Constraints on the velocity dispersion $\sigma_v$ and spatial extent $\sigma_\theta$ of the fluorescing iron K$\alpha$ emitting gas from a joint analysis of the {\it Chandra}/HETG, {\it Hitomi}/SXS and {\it XMM-Newton}/EPIC spectra.  Shown here are the 68\%, 90\%, 95\%, 99\%, and 99.9\% enclosed probability contours resulting from the MCMC analysis described in Section~\ref{reflectionmcmc}.  Also shown is the velocity-radius relation relevant for gravitational motions, specifically circular motion (red-dashed line) and radial infall (red-dotted line), assuming a stellar velocity dispersion of $\sigma_*=259\kmps$ \citep{riffel:20a} and central supermassive black hole of mass $M=1\times 10^9\Msun$.}
\label{fig:ironextent}
\end{figure}

Table~\ref{tab:jointfitting} lists the best fitting parameters (minimizing CSTAT) and the single-parameter error bounds resulting from the MCMC chain.  The ICM component as seen by {\it Hitomi} agrees well with the findings of \cite{hitomi:16a} and \cite{hitomi:18a}.  Since the {\it XMM-Newton}/EPIC spectra are extracted from a smaller region than for {\it Hitomi} and are subject to background subtraction that includes ICM emission, the residual ICM component in the EPIC spectra has a significantly lower normalization and is cooler (reflecting the fact that the coolest plasma reside  at the center of the system).

Marked changes are seen in the AGN component between the 2001 and 2016/2017 epochs.  We find an AGN 2--10\,keV flux that increases from $\sim 7\times 10^{-12}\ergpcmsqps$ in 2001 \citep[close to its historical minimum, ][]{fabian:15b} up to $\sim 2.5\times 10^{-11}\ergpcmsqps$ in 2016 and 2017. Concurrent with the historical brightening, we find that the AGN continuum becomes softer.  There is also a marked decrease in both the covering fraction and column density of the cold absorption from the 2001 to 2016/2017 epochs.  

The main motivation of this particular modelling is to obtain constraints on the velocity broadening and spatial extent of the iron-K$\alpha$ fluorescent line emission. Figure~\ref{fig:ironextent} shows confidence contours on the velocity width and spatial extent of the fluorescent iron line emitting gas.  The best-fitting model and 68\% enclosed probability contour suggest spatial broadening of the line emitting region at the 1\arcsec-scale.  However, the 90\% enclosed probability contour reaches down to the smallest extents probed ($0.01\arcsec$) and so in this case we do not claim the significant detection of extension.  Thus our result is not in strong tension with \cite{miller:17a} who used the novel technique of X-ray ``lucky imaging'' to constrain the extent of the iron-K$\alpha$ emission region to be $<0.3\arcsec$. 

This joint fit constrains the velocity width of the iron-K$\alpha$ line to be 310--963\kmps (90\% credible interval), translating into a full-width at half-maximum (FWMH) of 760--2300\kmps (90\% credible interval).  Interestingly, as evident from Fig.~\ref{fig:ironextent}, this analysis {\it disfavours} models with small spatial extent ($<50\pc$) and large velocity width ($\sigma_v>600\kmps$); this is a manifestation of the tension between the HETG and SXS line normalizations discussed above.  If we suppose that the dynamics of the iron-K$\alpha$ emitting material is gravitational, we can relate the distance from the black hole and the expected line-of-sight velocity dispersion $\sigma_g$,
\begin{equation}
\sigma_{\rm g}^2=\frac{f_v GM_{\rm BH}}{R}+\sigma_*^2
\end{equation}
where $\sigma_*$ is the velocity dispersion of the galaxy potential (approximated as an isothermal sphere).  Here, the factor $f_v=\sin i\approx 1/\sqrt{2}$ for circular motions (where $i$ is the viewing inclination that we take to be $i\sim 45^\circ$), and $f_v=2$ for radial infall.  These two gravitational velocities are shown on Fig.~\ref{fig:ironextent} (dotted and dashed red line).  We see that there is a tension between our measured line velocity and the expected gravitational motions if the fluorescing matter is closer than $R\sim 10\pc$.  

\section{Discussion and conclusions}\label{discussion}

The deep (490\ks) {\it Chandra}/HETG exposure presented in this work provides the most detailed view of the X-ray spectrum of NGC~1275 to date and further adds to the picture of a complex circumnuclear environment around this BCG-AGN.  
To recap our new findings:
\begin{enumerate}
\item The dominant component of the X-ray emission is well-approximated by an unabsorbed power-law.  We can clearly reject the hypothesis that the entire X-ray source is absorbed by the molecular gas column of $N_{H_2}\approx 2\times 10^{22}\pcmsq$ seen by ALMA towards the parsec-scale jets as reported by \cite{nagai:19a}.   However, we do find evidence that during our 2017 {\it Chandra} campaign approximately $15-20\%$ of the X-ray emission is absorbed by a cold (atomic and/or molecular) column density of $N_H\sim 8\times 10^{22}\pcmsq$, suggesting the possibility that some fraction of the X-ray emission does experience absorption by this molecular gas.   Applying the partial covering model to archival {\it XMM-Newton} EPIC data, we infer that both the covering fraction and column density of the absorber were higher in the 2001 epoch, a time when the overall X-ray luminosity was only 35\% of its 2017 level.  
\item A rigorous search that accounts for the look-elsewhere effect reveals no evidence for photoionized absorption in our HETG X-ray spectrum.  While a blind analysis of the high-resolution X-ray spectrum finds tentative detections several high ionization absorption/emission lines, the lack of a common velocity shift precludes them arising from a single absorber and we conclude that they are likely statistical fluctuations.  We rule out the presence of absorbers with ionization-parameters/column-densities commonly seen in Seyfert nuclei.
\item We detect the 6.4\keV iron-K$\alpha$ fluorescent line from cold circumnuclear gas that has been previously studied most recently by {\it Hitomi}.  Noting modest tensions between the iron-K$\alpha$ normalzations and/or widths between the {\it Chandra}/HETG, {\it Hitomi}/SXS, and {\it XMM-Newton}/EPIC spectra, we conduct a multi-satellite joint analysis that accounts for any anomalous broadening of the iron-K$\alpha$ line in the dispersed HETG that would result from spatial extension.  While we ultimately conclude that spatial broadening is not required at the 90\% confidence level, we believe that this combination of microcalorimeter data (which measures the true velocity width of a line) with high-quality grating data (with spectra that are convolved with the source spatial structure) will become importance once the {\it X-ray Imaging and Spectroscopy Mission (XRISM)} deploys.  
\end{enumerate}

To put these results into context, we must consider the parsec-scale structure of this AGN.   This AGN is remarkable in that it has been observed to go through very significant changes in its parsec scale structure on human timescales, as seen by high-resolution radio imaging of the associated radio source 3C~84. From the earliest days of Very Long Baseline Interferometry (VLBI), it was known that 3C~84 exhibits complex structure on milli-arcsecond (mas) scales \citep{paulinytoth:76a}.  At that epoch, the radio source was dominated by an inverted core and a second component located 1\,mas (0.4\pc) away on a position angle (PA) of 210$^\circ$ \citep{unwin:82a,readhead:83a,Readhead:83b}.  On slightly larger scales of 10--15\,mas (4--6\pc), fainter VLBI knots traced out a probably jet channel directly south (at a PA of 170--180$^\circ$) from the inverted core which appears to connect onto the kpc-scale jet with a PA of 170$^\circ$ \citep{pedlar:83a}.  VLBI monitoring revealed complex motions and flux changes of these jet subcomponents \citep{wright:88a,marr:89a,krichbaum:92a}.  The northern counter-jet was first reported in 22\,GHz VLBI image by \cite{vermeulen:94a} with an inverted spectrum that strongly suggested the influence of free-free absorption by an ionized screen with temperature $T\sim 10^4\K$ and column density $N_H\sim 10^{23}\pcmsq$ \citep{vermeulen:94a,walker:94a,walker:00a}.  The fact that the free-free screen absorbs the counter-jet but not the jet suggests that it be identified with the outer regions of the accretion disk.  

Both high-frequency (90\,GHz) radio monitoring and X-ray observations show that the AGN significantly faded from the time of these early VLBI studies and reached a minimum of activity around c2000 \citep{dutson:14a,fabian:15a}.  Since that time, we have been witnessing the onset of a remarkable new period of activity.  \cite{nagai:10a} analyzed 22\,GHz VLBI data to discover a new jet component (C3) propagating away from the core (C1) with PA of 180$^\circ$.  They initially suggest that the component was launched in 2005, although a later analysis of higher resolution 43\,GHz data showed that the C3 could already be distinguished from the C1 as early as 2003-November \citep{suzuki:12a}.   A picture emerged whereby C3 is the working surface of a new radio-lobe associated with restarted jet activity \citep{nagai:16a,kino:17a} and, c.2016, was approximately 3\,mas (1\,pc projected distance) directly south of the core C1.  This picture is supported by monitoring with the Korean VLBI Network (KVN).  This instrument shows weaker features in the jet that propagate from the core with apparent speeds ranging from 0.2--0.9c until they reach the vicinity of C3, after which they deflect and/or break apart, often coincident with a major $\gamma$-ray flare \citep{hodgson:21a}.  \cite{kino:18a} finds that the location of C3 skips eastwards by 0.4\,mas in 2015 August--September before resuming its southernly propagation, suggesting that the jet's working surface is pushing into a clumpy and dense medium; the density required to provide the necessary ram pressure, $n\sim 4\times 10^3\pcmcu-2\times 10^5\pcmcu$, suggests a clumpy molecular medium.   

Within this complex source, there are several possibilities for the origin of the nuclear X-rays including coronal emission from the inner accretion disk, the innermost regions of the jet (C1), the current working surface of the jet (C3), or the new jet-blown cocoon (seen in VLBI as the halo enveloping C1 and C3). Indeed, it would be natural for the observed X-rays to have multiple origins, the dominance of which can change over time. Monitoring by the {\it Fermi} Large Area Telescope (LAT) and the {\it Swift} X-ray Telescope (XRT) shows long term correlations between the nuclear 5--10\,keV X-ray flux and the 0.1--300\,GeV $\gamma$-ray flux which clearly has a jet/cocoon origin \citep{fukazawa:18a}.   However, given the small physical size of this source, this may simply be a reflection of the overall level of activity in the disk/jet system and does not necessarily imply that the bulk of the X-rays originate from the jet.  Short timescale $\gamma$-ray/X-ray correlations that would unambiguously signal a jet-origin for X-rays are more complex; some week-timescale $\gamma$-ray events show correlated X-ray variability whereas others do not \citep{fukazawa:18a,imazato:21a}.

In this context, our finding that the X-ray source is partially covered by a cold absorber fits naturally with the picture of a composite source embedded in a clumpy molecular medium. What is less clear is the precise identification of the absorbed and unabsorbed components.   Comparing the spectral fits to the 2001 {\it XMM-Newton} and 2017 {\it Chandra} datasets, we find that the factor of three increase in X-ray luminosity must be dominated by the unabsorbed component.   At the same time, the spectrum undergoes a significant softening, with the photon index increasing from $\Gamma\approx 1.65$ to $\Gamma\approx 2.0$.  This may suggest a transition from a jet/cocoon (non-thermal inverse Compton) dominated to a disk-corona (thermal Comptonization) dominated X-ray spectrum.   

Following this scenario, we infer that 80--85\% of the 2017-emission originates from the inner disk and is viewed along an unabsorbed line-of-sight, with the remaining 15--20\% coming from the parsec-scale jet structures including C3 absorbed by a column density $N_H\sim 10^{23}\pcmsq$ of molecular gas.  While the spectral models presented in this paper have formally assumed equal photon indices for the absorbed and unabsorbed components, we have verified that the fit to the 2017-HETG data is relatively insensitive to a decoupled photon index for the absorbed component (with allowable values in the range $\Gamma=1.1-2.5$ at 90\% confidence, safely bracketing the value of $\Gamma=1.65$).   This is not a unique conclusion, however --- fresh-injection of relativistic particles associated with jet shocks can create X-ray synchrotron components with $\Gamma\approx 2$ spectrum \citep{fukazawa:18a}.  Significantly more comparison of both short-and-long term X-ray variability with radio- and $\gamma$-ray variability will be needed to fully disentangle this composite source.

These results may be more general applicability to BCG AGN.  \cite{russell:13a} examine X-ray absorption in a sample of BCG-AGN, finding that 9/25 objects have significant columns ($N_H>10^{22}\pcmsq$) whereas 8/25 have no detectable absorption (with limits of $N_H<10^{21}\pcmsq$).  Extrapolating our findings for NGC~1275, this diversity in the broader population may result from the X-ray source in BCG-AGN being disk/jet composites with parsec-scale jets that are frustrated by dense molecular gas.  A natural consequence would be common mismatches between molecular line absorption and X-ray absorption.  Assuming that the remarkable decadal time variability of NGC~1275 is not unusual (applying Copernican reasoning), we may also expect to see year-to-decade timescale variability in the measurable X-ray absorption of other BCG-AGN as the dominance of the (absorbed) jet component waxes and wanes. 

\section*{Acknowledgements}

We thank the Chandra Science Center, and especially Hermann Marshall, for advice and guidance in the execution of these observations.  C.S.R. thanks the STFC for support under the New Applicant grant ST/R000867/1 and Consolidated Grant ST/S000623/1, as well as the European Research Council (ERC) for support under the European Union's Horizon 2020 research and innovation programme (grant 834203).  
R.S. and S.V. acknowledge support from NASA under the Chandra Guest Observer Program (grants G08-19088X and G09-20119X).

\section*{Data Availability}

 All of the raw data used in this work are available via the public data archives, specifically the Chandra Data Archive ({\it Chandra}/HETG), the NASA-GSFC High Energy Astrophysics Science Archive ({\it Hitomi}) and the {\it XMM-Newton} Science Archive ({\it XMM-Newton}).  Reduced data products and analysis scripts are available upon written request to the first author. 









\bsp	
\label{lastpage}
\end{document}